\documentclass[fleqn,11pt]{article}

\usepackage{amsmath,amsfonts,amssymb}
\usepackage{color}
\usepackage{geometry}
\newtheorem{theorem}{Theorem}
\newtheorem{lemma}[theorem]{Lemma}

\newtheorem{definition}{Definition}

\newcommand{\R}{\mathbb{R}}

\newcommand{\N}{\mathbb{N}}

\newcommand{\cqfd}
{%
\mbox{}%
\nolinebreak%
\hfill%
\rule{2mm}{2mm}%
\newline
\newline
}

\title{On the Cauchy problem with large data for a space-dependent Boltzmann-Nordheim boson equation.}
\author{Leif ARKERYD and Anne NOURI\\
\\Mathematical Sciences, 41296 G\"oteborg, Sweden,\\
arkeryd@chalmers.se\\
Aix-Marseille University, CNRS, Centrale Marseille, I2M UMR 7373, 13453 Marseille, France,\\
anne.nouri@univ-amu.fr}

\date{}

\begin{document}

\maketitle

{\noindent \bf Abstract.}\hspace{0.1in}
  This paper studies a Boltzmann Nordheim equation in a slab with two-dimensional velocity space and pseudo-Maxwellian forces. Strong  solutions are obtained for the Cauchy problem with large initial data in an $L^1\cap L^\infty$ setting. The main results are existence, uniqueness and stability of solutions conserving mass, momentum and energy that explode in $L^\infty $ if they are only local in time. The solutions are obtained as limits of solutions to corresponding anyon equations.

\footnotetext[2]{Key words; bosonic Boltzmann-Nordheim equation, low temperature kinetic theory, quantum Boltzmann equation.}
%
%
%
% FIRST SECTION: INTRODUCTION AND MAIN RESULT
%
\section{Introduction and main result.}
In a previous paper \cite{AN1}, we have studied the Cauchy problem for a space-dependent anyon Boltzmann equation,
\begin{equation}\label{f-alpha}
\partial _tf(t,x,v)+v_1\partial _xf(t,x,v)= Q_{\alpha }(f)(t,x,v),
\quad f(0,x ,v )= f_0(x,v)
,\hspace*{0.02in} {(t,x)\in \R _+\times [ 0,1] ,\hspace*{0.02in} v=(v_1,v_2)\in \R ^2.}
\end{equation}
The collision operator $Q_{\alpha }$ in \cite{AN1} depends on a parameter $\alpha \in ] 0,1[ $ and is given by 
\begin{eqnarray*}
Q_\alpha (f)(v)= \int_{I\! \!R^2 \times S^{1}}B(|v-v_*|,n)
 [f^\prime f^\prime _*F_\alpha(f)F_\alpha(f_*)-ff_*F_\alpha(f^\prime )F_\alpha(f^\prime _*)] dv_*dn,\hspace{.1cm}
\end{eqnarray*}
with the kernel B of Maxwellian type, $f^\prime $, $f^\prime _*$, $f$, $f_*$ the values of $f$ at $v^\prime $, $v^\prime _*$, $v$ and $v_*$ respectively, where
\begin{align*}
v^\prime = v-(v-v_*, n)n ,\quad v^\prime _*= v_*+(v-v_*, n)n \, ,
\end{align*}
and the filling factor $F_\alpha$
\begin{eqnarray*}
F_\alpha(f)= (1-\alpha f)^{\alpha}(1+(1-\alpha)f)^{1-\alpha}\, .
\end{eqnarray*}
Anyons are (quasi)particles that exist in one and two-dimensions besides fermions and bosons. The exchange of two identical anyons may cause a phase shift  different from $\pi $ (fermions) and $2\pi $ (bosons). In \cite{AN1}, also the limiting case $\alpha=1$ is discussed, a Boltzmann-Nordheim (BN) equation \cite{N} for fermions.
In the present paper we shall consider the other limiting case,  $\alpha=0$, which is a BN equation for bosons. \\
\\
For the bosonic BN equation general existence results were first obtained by X. Lu in \cite{Lu1} in the space-homogeneous isotropic boson large data case. It was followed by a number of interesting studies in the same isotropic setting, by X. Lu \cite{Lu2,Lu3,Lu4}, and by M. Escobedo and J.L. Vel\'azquez \cite{EV2,EV}. Results with the isotropy assumption removed, were recently obtained by M. Briant and A. Einav \cite{BE}. Finally a space-dependent case close to equilibrium has been studied by G. Royat in \cite{R}.\\
The papers \cite{Lu1,Lu2,Lu3,Lu4} by Lu, study the isotropic, space-homogeneous BN equation both for Cauchy data leading to mass and energy conservation, and for data leading to mass loss when time tends to infinity. Escobedo and Vel\'asquez in \cite{EV2,EV}, again in the isotropic space-homogeneous case, study initial data leading to concentration phenomena and blow-up in finite time of the $L^\infty$-norm of the solutions. The paper \cite{BE}  by Briant and Einav removes the isotropy restriction and obtain in polynomially weighted spaces of $L^1\cap L^\infty$ type, existence and uniqueness on a time interval $[0,T_0)$. In \cite{BE} either $T_0=\infty$, or for finite $T_0$ the $L^\infty$-norm of the solution tends to infinity, when time tends to $T_0$. Finally the paper \cite{R} considers the space-dependent problem, for a particular setting close to equilibrium, and proves well-posedness and convergence to equilibrium.
\\
The present paper studies a space-dependent, large data problem for the BN equation. The analysis is based on the anyon results in \cite{AN1}, which are
restricted to a slab set-up, since the proofs in \cite{AN1} use an estimate for the Bony functional only valid in one space dimension.
Due to the filling factor $F_\alpha (f)$, those proofs also in an essential way depend on the two-dimensional velocity frame. By a limiting procedure relying on the anyon case when $\alpha\rightarrow 0$, well-posedness and conservation laws are obtained in the present paper for the BN problen.\\
\setcounter{theorem}{0}
With
\begin{eqnarray*}
\cos \hspace*{0.02in} \theta = n \cdot \frac{v-v_*}{|v-v_*|}\, ,
\end{eqnarray*}
the kernel $B(|v-v_*|,n)$ will from now on be written $B(|v-v_*|, \theta )$ and assumed measurable with
\begin{equation}\label{hyp1-B}
0\leq B\leq B_0,
\end{equation}
for some $B_0>0$. It is also assumed for some $\gamma, \gamma',{c_B}>0$, that
\begin{equation}\label{hyp2-B}
B(|v-v_*|, \theta )=0 \hspace*{0.05in}\text{for}\hspace*{0.05in}   |\cos \hspace*{0.02in}\theta |<\gamma',\quad
\text{for}\hspace*{0.05in}  1-|\cos \hspace*{0.02in}\theta|<\gamma',\quad  \text{and for   } |v-v_*|< \gamma,
\end{equation}
 and that
 \begin{equation}\label{hyp3-B}
\int B(|v-v_*|, \theta )d\theta \geq c_B>0\quad  \text{for   }|v-v_*|\geq \gamma .
\end{equation}
These strong cut-off conditions on $B$ are made for mathematical reasons and assumed throughout the paper. For a more general discussion of cut-offs in the collision kernel $B$, see \cite{Lu2}. Notice that contrary to the classical Boltzmann operator where rigorous derivations of $B$ from various potentials have been made, little is known about collision kernels in quantum kinetic theory (cf \cite{V}).\\
With $v_1$ denoting the component of $v$ in the $x$-direction, the initial value problem for the Boltzmann Nordheim equation in a periodic in space setting is
\begin{equation}\label{f}
\partial _tf(t,x,v)+v_1\partial _xf(t,x,v)= Q(f)(t,x,v),
\end{equation}
where
\begin{equation}\label{Q}
Q(f)(v)= \int_{I\! \!R^2 \times [0,\pi ]}B(|v-v_*|,\theta)
 [f^\prime f^\prime _*F(f)F(f_*)-ff_*F(f^\prime )F(f^\prime _*)] dv_*d\theta ,\hspace{.1cm}
\end{equation}
and
\begin{equation}\label{F}
F(f)= 1+f.
\end{equation}
Denote by
\begin{equation}\label{f-sharp}
f^{\sharp }(t,x,v)= f(t,x+tv_1,v)\quad (t,x,v)\in \R _+\times [ 0,1] \times \R ^2.
\end{equation}
Strong solutions to the Boltzmann Nordheim paper are considered in the following sense.
\begin{definition}\label{strong-solution}
$f$ is a strong solution to (\ref{f}) on the time interval $I$ if
\begin{eqnarray*}
f\in\mathcal{C}^1(I;L^1([0,1]\times\R^2)),
\end{eqnarray*}
and
\begin{equation}\label{eq-along-characteristics}
\frac{d}{dt}f^{\sharp }= \big( Q(f)\big) ^{\sharp },\quad \text{on   } I\times [ 0,1] \times \R ^2.
\end{equation}
\end{definition}
The main result of this paper is the following.
\begin{theorem}\label{main-theorem}
Assume (\ref{hyp1-B})-(\ref{hyp2-B})-(\ref{hyp3-B}). Let $f_0\in L^\infty([0,1]\times\R^2)$ and satisfy
\begin{equation}\label{hyp-f0}
(1+|v|^2)f_0(x,v) \in L^1([ 0,1] \times \R ^2), \hspace{.1cm}
\int \sup_{x\in[0,1]} f_0(x,v)dv=c_0 <\infty,\hspace{.1cm}
\inf_{x\in[0,1]}f_0(x,v)>0,\hspace*{0.1cm}\text{a.a.}v\in \R ^2 .\hspace{.5cm}
\end{equation}
There exist a time $T_\infty >0$ and a strong solution $f$ to (\ref{f}) on $[0,T_\infty)$ with initial value $f_0$. \\
For $0<T<T_\infty$, it holds
\begin{equation}\label{regularity-f}
f^\sharp \in \mathcal{C}^1([0,T_\infty);L^1([0,1]\times\R^2))\cap L^\infty ([0,T]\times[0,1]\times \R^2).
\end{equation}
If $T_\infty <+\infty $ then
\begin{equation}\label{explosion}
\lim _{t\rightarrow T_\infty }\parallel f(t,\cdot ,\cdot )\parallel _{L^\infty ([ 0,1] \times \R ^2)}= +\infty .
\end{equation}
The solution is unique, depends continuously in $L^1$ on the initial value $f_0$, and conserves mass, momentum, and energy.
\end{theorem}
\underline{\bf Remark.}\\
A finite $T_\infty$  may not correspond to a condensation. In the isotropic space-homogeneous case considered in \cite{EV2,EV}, additional assumptions on the concentration of the initial value are considered in order to obtain condensation. \\
\\
{The paper is organized as follows.} In the following section, solutions $f_\alpha $ to the Cauchy problem for the anyon Boltzmann equation in the above setting are recalled, and their Bony functionals are uniformly controlled with respect to $\alpha $. In Section 3 the mass density of $f_\alpha $ is studied with respect to uniform control in $\alpha $. Theorem \ref{main-theorem} is proven in Section 4 except for the conservations of mass, momentum and energy that are proven in Section 5.
\hspace*{0.1in}\\
%
%
%  SECTION 2 PRELIMINARIES ON ANYONS AND THE BONY FUNCTIONAL
%
%
\section{Preliminaries on anyons and the Bony functional.}
\setcounter{equation}{0}
\setcounter{theorem}{0}
The Cauchy problem for a space-dependent anyon Boltzmann equation in a slab was studied in \cite{AN1}. That paper will be the starting point for the proof of Theorem \ref{main-theorem}, so we recall the main results from \cite{AN1}.
\begin{theorem}
\hspace*{0.1in}\\
Assume (\ref{hyp1-B})-(\ref{hyp2-B})-(\ref{hyp3-B}). Let the initial value $f_0$ be a measurable function on $[0,1]\times\R^2$ with values in $]0,\frac{1}{\alpha }] $, and satisfying (\ref{hyp-f0}).
For every $\alpha \in ] 0,1[ $, there exists a strong solution
$f_\alpha $ of (\ref{f-alpha}) with
\begin{align*}
f_\alpha ^\sharp \in \mathcal{C}^1([0,\infty [;L^1([0,1]\times\R^2)), \quad \quad 0<f_\alpha (t,\cdot ,\cdot )<\frac{1}{\alpha} \quad \text{for   }t>0,
\end{align*}
and
\begin{equation}\label{mass-density-f-alpha}
\int \sup _{(s,x)\in [0, t] \times [0,1] }f_\alpha ^\sharp (s,x,v)dv\leq c_\alpha(t) ,
\end{equation}
for some function  $c_\alpha(t) >0$  only depending on mass and energy. There is $t_m>0$ such that for any $T>t_m$, there is $\eta_T>0$ so that
\begin{eqnarray*}
f_\alpha (t,\cdot ,\cdot )\leq \frac{1}{\alpha}-\eta_T,\quad   t\in [t_m, T] .
\end{eqnarray*}
The solution is unique and {depends continuously in $\mathcal{C}([0,T];L^1([0,1]\times\R^2))$ on the initial $L^1$-datum.} It conserves mass, momentum and energy.
\end{theorem}
\[\]
The conditions $f_0\in L^\infty ([0,1] \times \R ^2)$ and (\ref{hyp-f0}) are assumed throughout the paper. \\
\\
{To obtain Theorem 1.1 for the boson BN equation from the anyon results, we start from a fixed initial value $f_0$ bounded by $2^L$ with $L\in \N$. We shall prove that there is
a time $T>0$ independent of $0<\alpha< 2^{-L-1}$, so that the solutions are bounded by $2^{L+1}$ on $[0,T]$. For that, some lemmas from the anyon paper are sharpened to obtain control in terms of only mass, energy and L.  We then prove that the limit $f$ of the solutions $f_\alpha$ when $\alpha\rightarrow 0$  solves the corresponding bosonic  BN problem.
Iterating the result from T on, it follows that $f$ exists up to the first time $T_\infty$ when $ \lim_{t\rightarrow T_\infty}  \parallel f_\alpha (t,\cdot ,\cdot )\parallel _{L^{\infty }([0,1] \times \R ^2)} = \infty$.\\
\\
\\
%
% Lemma 2.2
%
We observe that
\begin{lemma}\label{T-dependent-on-alpha}
\hspace{.1cm}\\
Given $f_0\leq2^L$ and satisfying (\ref{hyp-f0}), there is for each $\alpha \in ]0,2^{-L-1}[$ a time $T_\alpha >0$ so that the solution $f_\alpha$ to (\ref{f-alpha}) is bounded by $2^{L+1}$ on $[0,T_\alpha ]$.
\end{lemma}
\underline{Proof of Lemma \ref{T-dependent-on-alpha}.}\\
Split the Boltzmann anyon operator $Q_\alpha $ into $Q_\alpha = Q_\alpha ^+-Q_\alpha ^-$, where the gain (resp. loss) term $Q_\alpha ^+$ (resp. $Q_\alpha ^-$) is defined by
\begin{equation}\label{gain-loss}
Q_\alpha ^+(f)(v)= \int Bf^\prime f^\prime _*F_\alpha (f)F_\alpha (f_*)dv_*d\theta \quad (resp. \hspace*{0.05in}Q_\alpha ^-(f)(v)= \int Bff_*F_\alpha (f^\prime )F_\alpha (f^\prime _*)dv_*d\theta ).
\end{equation}
The solution $f_\alpha $ to (\ref{f-alpha}) satisfies
\begin{eqnarray*}
 f_\alpha ^{\sharp }(t,x,v)= f_0(x,v)+\int_0^tQ_\alpha(f_\alpha )(s,x+sv_1,v)ds\leq
f_0(x,v)+\int_0^tQ^{+}_\alpha (f_\alpha )(s,x+sv_1,v)ds.
 \end{eqnarray*}
Hence
\begin{align}\label{control-by-gain}
&\sup_{s\leq t}f_\alpha ^{\sharp} (s,x,v)
\leq  f_0(x,v)+\int_0^tQ_{\alpha }^{+}(f_\alpha )(s,x+sv_1,v)ds\\
&= f_0(x,v)+\int_0^t\int Bf_\alpha (s,x+sv_1,v^\prime )f_\alpha (s,x+sv_1,v^\prime _*)
F_\alpha (f_\alpha )(s,x+sv_1,v)F_\alpha (f_\alpha )(s,x+sv_1,v_*)dv_*d\theta ds\nonumber \\
&\leq 2^L+\frac{B_0}{\alpha }\Big( \frac{1}{\alpha }-1\Big) ^{2(1-2\alpha )}\int _0^t\int f_\alpha (s,x+sv_1,v^\prime )dv_*d\theta ds, \nonumber
\end{align}
since the maximum of $F_\alpha $ on $[ 0,\frac{1}{\alpha }]$ is $(\frac{1}{\alpha }-1)^{1-2\alpha }$ for $\alpha \in ] 0,\frac{1}{2}[ $. With the angular cut-off (2.2), $v_*  \rightarrow v^\prime $ is a change of variables. Using it and (\ref{mass-density-f-alpha}) for $t\leq 1$ leads to
\begin{align*}
\sup_{s\leq t,x}f_{\alpha }^{\sharp} (s,x,v)&\leq 2^L+ c\frac{B_0c_\alpha (1)}{\alpha }\Big( \frac{1}{\alpha }-1\Big) ^{2(1-2\alpha )}t\\
&\leq 2^{L+1}\hspace*{1.7in}\text{for }t\leq \min \{1, \frac{2^L\alpha ^{3-4\alpha }(1-\alpha )^{2(2\alpha -1)}}{cB_0c_\alpha (1)}\} \, .
\end{align*}
The lemma follows.\cqfd\\
The estimate of the Bony functional
\begin{eqnarray*}
\bar{B}_\alpha (t):= \int_0^1\int |v-v_*|^2 Bf_\alpha f_{\alpha *}F_\alpha (f^\prime _\alpha )F_\alpha (f^\prime _{\alpha *})dvdv_*d\theta dx,\quad t\geq 0,
\end{eqnarray*}
from the proof of Theorem 2.1 for $f_\alpha\leq 2^{L+1}$ , can be sharpened.
%
%
% Lemma 2.3
%
\begin{lemma}\label{Bony}
\hspace*{0.1in}\\
For $\alpha \leq 2^{-L-1}$ and $T>0$ such that  $f_\alpha (t)\leq2^{L+1}$ for $0\leq t\leq T$, it holds
\begin{eqnarray*}
\int_0^{T }\bar{B}_\alpha (t)dt\leq c'_0(1+T)\, ,
\end{eqnarray*}
{with $c'_0$ independent of $T$ and $\alpha $, and only depending on $\int f_0(x,v)dxdv$, $\int |v|^2f_0(x,v)dxdv$ and $L$.}
\end{lemma}
\underline{Proof of Lemma \ref{Bony}.} \\
Denote $f_\alpha $ by $f$ for simplicity. The proof {is} an extension of the classical one (cf \cite{B}, \cite{CI}), together with the control of the filling factor $F_\alpha $ when $v\in \R ^2$, as follows. \\
The integral over time of the momentum $\int v_1f(t,0,v)dv$ (resp. the momentum flux
\\
$\int v_1^2f(t,0,v)dv$ ) is first controlled. Let $\beta \in C^1([ 0,1] )$ be such that $\beta (0)= -1$ and $\beta (1)= 1$. Multiply (\ref{f-alpha}) by $\beta (x)$ (resp. $v_1\beta (x)$ ) and integrate over $[ 0,t] \times [ 0,1] \times \R ^2$. It gives
\[ \begin{aligned}
\int _0^t\int v_1f(\tau ,0,v)dvd\tau = \frac{1}{2}\big( \int \beta (x)f_0(x,v)dxdv&-\int \beta (x)f(t,x,v)dxdv\\
&+\int _0^t\int \beta ^\prime (x)v_1f(\tau ,x,v)dxdvd\tau\big) ,
\end{aligned}\]
\Big( resp.
\[ \begin{aligned}
\int _0^t\int v_1^2f(\tau ,0,v)dvd\tau = \frac{1}{2}\big( \int \beta (x)v_1f_0(x,v)dxdv&-\int \beta (x)v_1f(t,x,v)dxdv\\
&+\int _0^t\int \beta ^\prime (x)v_1^2f(\tau ,x,v)dxdvd\tau\big) \Big) .
\end{aligned}\]
Consequently, using the conservation of mass and energy of $f$,
\begin{align}\label{bony-1}
\lvert \int _0^t\int v_1f(\tau ,0,v)dvd\tau \rvert +\int _0^t\int v_1^2f(\tau ,0,v)dvd\tau \leq c(1+t).
\end{align}
Here $c$ is of magnitude of mass plus energy uniformly in $\alpha$.
Let
\begin{eqnarray*}
\mathcal{I}(t)= \int _{x<y}(v_1-v_{*1})f(t,x,v)f(t,y,v_*)dxdydvdv_*.
\end{eqnarray*}
It results from
\begin{eqnarray*}
\mathcal{I}'(t)= -\int (v_1-v_{*1})^2f(t,x,v)f(t,x,v_*)dxdvdv_*+2\int v_{*1}(v_{*1}-v_1)f(t,0,v_*)f(t,x,v)dxdvdv_*,
\end{eqnarray*}
and the conservations of the mass, momentum and energy of $f$ that
\begin{align}\label{bony-2}
&\int _0^t \int_0^1 \int (v_1-v_{*1})^2 f(s,x,v)f(s,x,v_*)dvdv_*dxds\nonumber \\
&\leq 2\int f_0(x,v)dxdv\int \lvert v_1\rvert f_0(x,v)dv+ 2\int f(t,x,v)dxdv\int \lvert v_1\rvert f(t,x,v)dxdv\nonumber \\
&+2\int _0^t\int v_{*1}(v_{*1}-v_1)f(\tau ,0,v_*)f(\tau ,x,v)dxdvdv_*d\tau \nonumber \\
&\leq 2\int f_0(x,v)dxdv\int (1+\lvert v\rvert ^2)f_0(x,v)dv+ 2\int f(t,x,v)dxdv\int (1+\lvert v\rvert ^2) f(t,x,v)dxdv\nonumber \\
&+2\int _0^t(\int v_{*1}^2f(\tau ,0,v_*)dv_*)d\tau\int f_0(x,v)dxdv-2\int _0^t(\int v_{*1}f(\tau ,0,v_*)dv_*)d\tau\int  v_1f_0(x,v)dxdv\nonumber \\
&\leq c\Big( 1+\int _0^t\int v_1^2f(\tau ,0,v)dvd\tau +\lvert \int _0^t\int v_1f(\tau ,0,v)dvd\tau \rvert \Big) . \nonumber
\end{align}
And so, by (\ref{bony-1}),
\begin{equation}\label{bony-3}
\int _0^t \int_0^1 \int (v_1-v_{*1})^2 f(s,x,v)f(s,x,v_*)dvdv_*dxds\leq c(1+t).
\end{equation}
Denote by $u_1=\frac{\int v_1fdv}{\int fdv}$. Recalling (\ref{hyp1-B}) it holds
\begin{align}\label{bony-4}
\int_0^t\int_0^1 \int (v_1-u_1)^2 B{\chi}_jff_*&F_j(f^\prime )F_j(f^\prime _*)(s,x,v,v_*,\theta )dvdv_*d\theta dxds\nonumber \\
&\leq c\int_0^t  \int_0^1 \int (v_1-u_1)^2 ff_*(s,x,v,v_*)dvdv_* dxds\nonumber \\
&= \frac{c}{2}\int _0^t \int_0^1 \int (v_1-v_{*1})^2 ff_*(s,x,v,v_*)dvdv_*dxds\nonumber \\
&\leq c(1+t).
\end{align}
Here $c$ also contains $\sup F_\alpha (f')F_\alpha (f'_*)$ which is of magnitude bounded by $2^{2L}$. So $c$ is of magnitude $2^{2L}$(mass+energy) and uniformly in $\alpha$. Multiply equation (\ref{f-alpha}) for $f$  by $v_1^2$, integrate and use that $\int v_1^2Q_\alpha (f)dv= \int (v_1-u_1)^2Q_\alpha (f)dv$ and (\ref{bony-4}). It results
\[ \begin{aligned}
&\int _0^t\int (v_1-u_1)^2Bf^\prime f^\prime _*F_\alpha (f)F_\alpha (f_*)dvdv_*d\theta dxds\\
&= \int v_1^2f(t,x,v)dxdv-\int v_1^2f_0(x,v)dxdv+\int _0^t\int (v_1-u_1)^2Bff_*F_\alpha (f^\prime )F_\alpha (f^\prime _*)dxdvdv_*d\theta ds\\
&<c_0(1+t),
\end{aligned}\]
where $c_0$ is a constant {of magnitude $2^{2L}$(mass+energy)}.
\hspace{1cm}\\
After  a change of variables the left hand side can be written
\[ \begin{aligned}
&\int _0^t\int (v'_1-u_1)^2Bff_*F_\alpha (f^\prime )F_\alpha (f^\prime _*)dvdv_*d\theta dxds\\
&= \int _0^t\int (c_1-n_1[(v-v_*)\cdot n])^2Bff_*F_\alpha (f^\prime )F_\alpha (f^\prime _*)dvdv_*d\theta dxds,
\end{aligned}\]
where $c_1=v_1-u_1$. And so,
\[ \begin{aligned}
&\int _0^t\int n_1^2[(v-v_*)\cdot n])^2Bff_*F_\alpha (f^\prime )F_\alpha (f^\prime _*)dvdv_*d\theta dxds\\
&\leq c_0(1+t)+2\int _0^t\int c_1n_1[(v-v_*)\cdot n]Bff_*F_\alpha (f^\prime )F_\alpha (f^\prime _*)dvdv_*d\theta dxds.
\end{aligned}\]
\\
The term containing $n_1^2[(v-v_*)\cdot n]^2$ is estimated from below.
When $n$ is replaced by an orthogonal (direct) unit vector $n_\perp $, $v^\prime $ and $v^\prime _*$ are shifted and the product $ff_*F_\alpha (f^\prime )F_\alpha (f^\prime _*)$ is unchanged. In $\R^2$ the ratio between the sum of the integrand factors $n_1^2[(v-v_*)\cdot n]^2+ n_{\perp 1}^2[(v-v_*)\cdot n_{\perp}]^2$ and $|v-v_*|^2$, is, outside of the angular cut-off {(\ref{hyp2-B})}, uniformly bounded from below by {$\gamma ^{\prime 2}$}. Indeed, if $\theta $ (resp. $\theta _1$) denotes the angle between $\frac{v-v_*}{ |v-v_*|}$ and $n$ (resp. the angle between $e_1$ and $n$, where $e_1$ is a unit vector in the $x$-direction), 
\[ \begin{aligned}
n_1^2[\frac{v-v_*}{ |v-v_*|}\cdot n]^2+ n_{\perp 1}^2[\frac{v-v_*}{ |v-v_*|}\cdot n_{\perp}]^2&= \cos ^2\theta _1\hspace*{0.02in}\cos ^2\theta +\sin ^2\theta _1\hspace*{0.02in}\sin ^2\theta \\
&\geq \gamma ^{\prime 2}\cos ^2\theta_1 +\gamma ^\prime (2-\gamma^\prime )\sin ^2\theta_1 \\
&\geq\gamma ^{\prime 2},\quad \gamma ^\prime <|\cos \theta | <1-\gamma ^\prime ,\quad \theta_1 \in [ 0,2\pi ] .
\end{aligned}\]
This is where the condition $v\in \R ^2$ is used.\\
\hspace*{0.1in}\\
That leads to the lower bound
\begin{eqnarray*}
\int_0^t\int n_1^2[(v-v_*)\cdot n]^2Bff_*F_\alpha (f^\prime )F_\alpha (f^\prime _*)dvdv_*d\theta dxds\\
\geq \frac{\gamma ^{\prime 2}}{2}\int _0^t \int |v-v_*|^2Bff_*F_\alpha (f^\prime )F_\alpha (f^\prime _*)dvdv_*d\theta dxds.
\end{eqnarray*}
And so,
\[ \begin{aligned}
&{\gamma ^{\prime 2}}\int _0^t \int |v-v_*|^2Bff_*F_\alpha (f^\prime )F_\alpha (f^\prime _*)dvdv_*d\theta dxds\\
&\leq 2c_0{(1+t)}+4\int _0^t\int (v_1-u_1)n_1[(v-v_*)\cdot n]Bff_*F_\alpha (f^\prime )F_\alpha (f^\prime _*)dvdv_*d\theta dxds\\
&\leq 2{c_0}{(1+t)}+4\int _0^t\int \Big( v_1(v_2-v_{*2})n_1n_2
\Big) Bff_*F_\alpha (f^\prime )F_\alpha (f^\prime _*)dvdv_*d\theta dxds,
\end{aligned}\]
since
\[ \begin{aligned}
\int &u_1(v_1-v_{*1})n_1^2Bff_*F_\alpha (f^\prime )F_\alpha (f^\prime _*)dvdv_*d\theta dx\\
&= \int u_1(v_2-v_{*2})n_1n_2Bff_*F_\alpha (f^\prime )F_\alpha (f^\prime _*)dvdv_*d\theta dx
= \hspace*{0.01in}0,
\end{aligned}\]
by an exchange of the variables $v$ and $v_*$. Moreover, exchanging first the variables $v$ and $v_*$,
\[ \begin{aligned}
2\int _0^t&\int v_1(v_2-v_{*2})n_1n_2Bff_*F_\alpha (f^\prime )F_\alpha (f^\prime _*)dvdv_*d\theta dxds\\
= &\int _0^t\int (v_1-v_{*1})(v_2-v_{*2})n_1n_2Bff_*F_\alpha (f^\prime )F_\alpha (f^\prime _*)dvdv_*d\theta dxds\\
\leq &\frac{8}{\gamma ^{\prime 2}}\int _0^t\int (v_1-v_{*1})^2n_1^2Bff_*F_\alpha (f^\prime )F_\alpha (f^\prime _*)dvdv_*d\theta dxds\\
&+\frac{{\gamma ^{\prime 2}}}{8}\int _0^t\int (v_2-v_{*2})^2n_2^2Bff_*F_\alpha (f^\prime )F_\alpha (f^\prime _*)dvdv_*d\theta dxds\\
\leq &\frac{8\pi c_0}{{\gamma ^{\prime 2}}}{(1+t)}+\frac{{\gamma ^{\prime 2}}}{8}\int _0^t\int (v_2-v_{*2})^2n_2^2Bff_*F_\alpha (f^\prime )F_\alpha (f^\prime _*)dvdv_*d\theta dxds.
\end{aligned}\]
It follows that
\begin{eqnarray*}
\int _0^t \int |v-v_*|^2Bff_*F_\alpha (f^\prime )F_\alpha (f^\prime _*)dvdv_*d\theta dxds\leq c'_0(1+t),
\end{eqnarray*}
with $c'_0$ {uniformly with respect to $\alpha$, of the same magnitude as $c_0$, only depending on $\int f_0(x,v)dxdv$, $\int \lvert v\rvert ^2f_0(x,v)dxdv$  and $L$}. This completes the proof of the lemma. \cqfd
\\
\\
%
%
% SECTION 3 CONTROL OF PHASE SPACE DENSITY
%
%
\section{Control of phase space density.}
\setcounter{equation}{0}
This section is devoted to obtaining a time $T>0$, such that 
\begin{eqnarray*}
\sup _{t\in [0,T],\hspace*{0.02in}x\in [ 0,1] }f_\alpha ^\sharp (t,x,v)\leq 2^{L+1},
\end{eqnarray*}
uniformly with respect to  $\alpha \in ]0,2^{-L-1}[ $ .
We start from the case of a fixed $\alpha\leq 2^{-L-1}$. Up to Lemma 3.3 the time interval when the solution does not exceed $2^{L+1}$, may be $\alpha$-dependent. Lemma \ref{T-independent-on-alpha} implies that this time interval can be chosen independent of $\alpha$. \\
\setcounter{theorem}{0}
\[\]
%
% Lemma 3.1
%
\begin{lemma}\label{integral-dxdv}
\hspace*{0.2in}\\
Given $T>0$ such that  $f_\alpha (t)\leq2^{L+1}$ for $0\leq t\leq T$, the solution $f_\alpha $ of (\ref{f-alpha}) satisfies
\begin{eqnarray*}
\int \sup_{t\in [ 0, T] }f_\alpha ^\sharp (t,x,v)dxdv<c^\prime _1+c^\prime _2T,\quad \alpha \in ] 0,2^{-L-1}[ ,
\end{eqnarray*}
{where $c'_1$ and $c_2'$ are independent of $T$ and $\alpha$, and only depend on $\int f_0(x,v)dxdv$, $\int |v|^2f_0(x,v)dxdv$ and $L$.}
%where $c^\prime _1$ and $c^\prime _2$ only depend on $T$, {$\int f_0(x,v)dxdv$ and $\int |v|^2f_0(x,v)dxdv$.}
\end{lemma}
\underline{Proof of Lemma \ref{integral-dxdv}.} \\
Denote $f_\alpha $ by $f$ for simplicity. By (\ref{control-by-gain}),
\begin{eqnarray*}
{\sup_{t\in [ 0, T] }f^\sharp (t,x,v)\leq  f_0(x,v)+\int_0^{T }Q_\alpha ^+(f)(t,x+tv_1,v)dt.}
\end{eqnarray*}
Integrating the previous inequality with respect to $(x,v)$ and using Lemma \ref{Bony}, gives
{\[ \begin{aligned}
\int \sup_{0\leq t\leq T}f^\sharp (t,x,v)dxdv
&\leq  \int f_0(x,v)dxdv+\int_0^{T }\int B\\
f(t,x+tv_1,v')f(t,x+tv_1,v'_*)&F_\alpha (f)(t,x+tv_1,v)F_\alpha (f)(t,x+tv_1,v_*)dvdv_*d\theta dxdt\\
&\leq  \int f_0(x,v)dxdv+\frac{1}{\gamma^2}\int_0^{T }\int B|v-v_*|^2
\\
f(t,x,v')f(t,x,v'_*)&F_\alpha (f)(t,x,v)F_\alpha (f)(t,x,v_*)dvdv_*d\theta dxdt\\
&\leq  \int f_0(x,v)dxdv+{\frac{c_0'(1+T)}{\gamma^2}}:= \frac{C_1+C_2T}{\gamma^2}
.\quad \quad \quad \quad \quad \cqfd
\end{aligned}\]
}
\[\]
%
% Lemma 3.2
%
\begin{lemma}\label{on-small-sets}
\hspace*{0.2in}\\
Given $T>0$ such that  $f(t)\leq2^{L+1}$ for $0\leq t\leq T$, and $\delta_1>0$, there exist $\delta_2>0$ and $t_0>0$ {independent of $T$ and $\alpha$ and only depending on
$\int f_0(x,v)dxdv$, $\int |v|^2f_0(x,v)dxdv$ and $L$}, such that
\begin{eqnarray*}
\sup _{x_0\in[0,1] }\int_{|x-x_0|<\delta_2} \hspace*{0.03in}{\sup_{t\leq s\leq t+t_0}}f_\alpha ^\sharp (s,x,v)dxdv<\delta_1,\quad \alpha \in ]0,2^{-L-1}[ ,\quad t\in [0,T] .
\end{eqnarray*}
\end{lemma}
\underline{Proof of Lemma \ref{on-small-sets}.} \\
Denote $f_\alpha $ by $f$ for simplicity. For $s\in [ t, t+t_0] $ it holds,
\[ \begin{aligned}
 f^\sharp (s,x,v)&=f^{\sharp}(t+t_0,x,v)-\int_{s}^{t+t_0}Q_\alpha (f)(\tau,x+\tau v_1,v)d\tau\\
&\leq  f^{\sharp}(t+t_0,x,v)+\int_{s}^{t+t_0}Q_\alpha ^-(f)(\tau,x+\tau v_1,v)d\tau.
\end{aligned}\]
And so
\begin{eqnarray*}
 \sup_{t\leq s\leq t+t_{0}}f^\sharp (s,x,v)
\leq  f^{\sharp}(t+t_0,x,v)+\int_{t}^{t+t_0}Q_\alpha ^-(f)(s,x+sv_1,v)ds.
\end{eqnarray*}
Integrating with respect to $(x,v)${, using Lemma \ref{Bony} and the bound {$2^{L+1}$} from above for $f$}, gives
\[ \begin{aligned}
&\int_{|x-x_0|<\delta_2} \sup_{t\leq s\leq t+t_0}f^\sharp (s,x,v)dxdv\\
&\leq  \int_{|x-x_0|<\delta_2} f^{\sharp}(t+t_0,x,v)dxdv\\
&+\int_{t}^{t+t_0}\int Bf^{\sharp}(s,x,v)f(s,x+sv_1,v_*)F_\alpha (f)(s,x+sv_1,v')
F_\alpha (f)(s,x+sv_1,v'_*)dvdv_*d\theta dxds\\
\end{aligned}\]
\[ \begin{aligned}
&\leq  \int_{|x-x_0|<\delta_2} f^{\sharp}(t+t_0,x,v)dxdv
+\frac{1}{\lambda^2}\int_{t}^{t+t_0}\int_{|v-v_*|\geq\lambda} B|v-v_*|^2
f^{\sharp}(s,x,v)f(s,x+sv_1,v_*)\\
&F_\alpha (f)(s,x+sv_1,v')F_\alpha (f)(s,x+sv_1,v'_*)dvdv_*d\theta dxds\\
&
{+c2^{2L}\int_{t}^{t+t_0}\int _{|v-v_*|<\lambda} B
f^{\sharp}(s,x,v)f(s,x+sv_1,v_*)dvdv_*
d\theta dxds}\\
&{\leq  \int_{|x-x_0|<\delta_2} f^{\sharp}(t+t_0,x,v)dxdv
+\frac{c_0'(1+t_0)}{\lambda^2}+c2^{3L}t_0 \lambda^2\int f_0(x,v)dxdv}\\
&{\leq \frac{1}{\Lambda^2}\int v^2f_0dxdv + c\delta_2 2^L
\Lambda^2+\frac{c_0'(1+t_0)}{\lambda^2}+c2^{3L}t_0 \lambda^2\int
f_0(x,v)dxdv.}
\end{aligned}\]
Depending on $\delta_1$, suitably choosing $\Lambda$ and then $\delta_2$, $\lambda$ and then $t_0$, the lemma follows.   \cqfd
{The previous lemmas imply for fixed $\alpha\leq2^{-L-1}$ {a bound} for the $v$-integral of $f_\alpha ^\#$ only depending on
$\int f_0(x,v)dxdv$, $\int |v|^2f_0(x,v)dxdv$ and $L$.}
%
%
% Lemma 3.3
%
\begin{lemma}\label{control-mass-density}
\hspace*{0.1in}\\
{With $T^\prime_\alpha $ defined as the maximum time for which $f_\alpha (t)\leq 2^{L+1}$, $t\in [0,T^\prime_\alpha]$, take $T_\alpha=\min\{1,T^\prime_\alpha \}$. \\
The solution $f_\alpha$ of (\ref{f-alpha}) satisfies
\begin{equation}\label{mass-density}
\int \sup_{(t,x)\in [0,T_\alpha[ \times [ 0,1] }f_\alpha^\sharp (t,x,v)dv\leq c_1 ,
\end{equation}
where $c_1$ is independent of $\alpha\leq 2^{-L-1}$ and only depends on $\int f_0(x,v)dxdv$, $\int |v|^2f_0(x,v)dxdv$ and $L$.}
\end{lemma}
\underline{Proof of Lemma \ref{control-mass-density}.} \\
Denote by $E(x)$ the integer part of $x\in\R$, $E(x)\leq x< E(x)+1$.\\
By (\ref{control-by-gain}),
\begin{align}
 &\sup_{s\leq t}f^\sharp (s,x,v)
\leq  f_0(x,v)+\int_0^tQ_\alpha ^+(f)(s,x+sv_1,v)ds\nonumber \\
&= f_0(x,v)
+\int_0^t\int Bf(s,x+sv_1,v')f(s,x+sv_1,v'_*)F_\alpha (f)(s,x+sv_1,v)F_\alpha (f)(s,x+sv_1,v_*)dv_*d\theta ds\nonumber\\
&\leq  f_0(x,v)+c{2^{2L}}A,
\end{align}
where
\begin{eqnarray*}
A= \int_0^t\int B
 \sup _{\tau \in [ 0,t]} f^{\#}(\tau ,x+s(v_1-v^\prime _1),v^\prime ) \sup _{\tau \in [ 0,t]}f^{\#}(\tau ,x+s(v_1-{v^\prime }_{*1}),v^\prime _*)dv_*d\theta ds.\quad
\end{eqnarray*}
For $\theta $ outside of the angular cutoff {(2.2)}, let $n$ be the unit vector in the direction $v-v'$, and $n_{\perp}$ the orthogonal
unit vector in the direction $v-v'_*$. With $e_1$ a unit vector in the $x$-direction,
\begin{eqnarray*}
\max(|n\cdot e_1|,|n_{\perp}\cdot e_1|)\geq \frac{1}{\sqrt{2}}.
\end{eqnarray*}
For $\delta _2>0$ that will be fixed later, split $A$ into $A_1+A_2+A_3+A_4$, where
\begin{eqnarray*}
A_1= \int_0^t\int _{|n\cdot e_1|\geq \frac{1}{\sqrt{2}},\hspace*{0.02in}t|v_1-v^\prime _1|>\delta _2}B\sup _{\tau \in [ 0,t]} f^{\#}(\tau ,x+s(v_1-v'_1),v') \sup _{\tau \in [ 0,t]}f^{\#}(\tau ,x+s(v_1-{v'}_{*1}),v'_*)dv_*d\theta ds,\\
A_2= \int_0^t\int _{|n\cdot e_1|\geq \frac{1}{\sqrt{2}},\hspace*{0.02in}t|v_1-v^\prime _1|<\delta _2}B\sup _{\tau \in [ 0,t]} f^{\#}(\tau ,x+s(v_1-v'_1),v') \sup _{\tau \in [ 0,t]}f^{\#}(\tau ,x+s(v_1-{v'}_{*1}),v'_*)dv_*d\theta ds,\\
A_3= \int_0^t\int _{|n_\perp \cdot e_1|\geq \frac{1}{\sqrt{2}},\hspace*{0.02in}t|v_1-v^\prime _1|>\delta _2}B\sup _{\tau \in [ 0,t]} f^{\#}(\tau ,x+s(v_1-v'_1),v') \sup _{\tau \in [ 0,t]}f^{\#}(\tau ,x+s(v_1-{v'}_{*1}),v'_*)dv_*d\theta ds,\\
A_4= \int_0^t\int _{|n_\perp \cdot e_1|\geq \frac{1}{\sqrt{2}},\hspace*{0.02in}t|v_1-v^\prime _1|<\delta _2}B\sup _{\tau \in [ 0,t]} f^{\#}(\tau ,x+s(v_1-v'_1),v') \sup _{\tau \in [ 0,t]}f^{\#}(\tau ,x+s(v_1-{v'}_{*1}),v'_*)dv_*d\theta ds.
\end{eqnarray*}
In $A_1$ and $A_2$, bound the factor $\sup _{\tau \in [ 0,t] }f^\sharp (\tau ,x+s(v_1-v^\prime _{*1}),v^\prime _*)$ by its supremum over $x\in [ 0,1] $, and make the change of variables
\begin{eqnarray*}
 s\rightarrow y= x+s(v_1-v^\prime _1).
\end{eqnarray*}
with Jacobian
\begin{eqnarray*}
\frac{Ds}{Dy}= \frac{1}{|v_1-v^\prime _1|}= \frac{1}{|v-v_*|\hspace*{0.03in}|(n,\frac{v-v_*}{|v-v_*|})|\hspace*{0.03in}|n_1|} \leq \frac{\sqrt{2}}{\gamma \gamma ^\prime }.
\end{eqnarray*}
It holds that
\begin{eqnarray*}
A_1\leq \int _{t|v_1-v^\prime _1|>\delta _2}\frac{B}{|v_1-v^\prime _1|}\Big( \int _{y\in (x,x+t(v_1-v^\prime _1))}\sup _{\tau \in [ 0,t]} f^{\#}(\tau ,y,v')dy\Big)  \sup _{(\tau ,X)\in [ 0,t]\times [ 0,1] }f^{\#}(\tau ,X,v'_*)dv_*d\theta ,
\end{eqnarray*}
and
\begin{eqnarray*}
A_2\leq \frac{\sqrt{2}}{\gamma \gamma ^\prime }\int _{|n\cdot e_1|\geq \frac{1}{\sqrt{2}},\hspace*{0.02in}t|v_1-v^\prime _1|<\delta _2}B\Big( \int _{|y-x|<\delta _2}\sup _{\tau \in [ 0,t]} f^{\#}(\tau ,y,v')dy\Big)  \sup _{(\tau ,X)\in [ 0,t]\times [ 0,1] }f^{\#}(\tau ,X,v'_*)dv_*d\theta .
\end{eqnarray*}
Then, performing the change of variables $(v,v_*,n )\rightarrow (v^\prime ,v^\prime _*,-n )$,
\[ \begin{aligned}
&\int \sup _{x\in [ 0,1] }A_1dv\\
&\leq \int _{t|v_1-v^\prime _1|>\delta _2}\frac{B}{|v_1-v^\prime _1|}\sup _{x\in [ 0,1] }\Big( \int _{y\in (x,x+t(v^\prime _1-v_1))}\sup _{\tau \in [ 0,t]} f^{\#}(\tau ,y,v)dy\Big)  \sup _{(\tau ,X)\in [ 0,t]\times [ 0,1] }f^{\#}(\tau ,X,v_*)dvdv_*d\theta ,
\end{aligned}\]
so that
\[ \begin{aligned}
&\int \sup _{x\in [ 0,1] }A_1dv\\
&\leq \int _{t|v_1-v^\prime _1|>\delta _2}\frac{B}{|v_1-v^\prime _1|}\sup _{x\in [ 0,1] }\Big( \int _{y\in (x,x+E(t(v^\prime _1-v_1)+1))}\sup _{\tau \in [ 0,t]} f^{\#}(\tau ,y,v)dy\Big)  \sup _{(\tau ,X)\in [ 0,t]\times [ 0,1] }f^{\#}(\tau ,X,v_*)dvdv_*d\theta \\
&= \int _{t|v_1-v^\prime _1|>\delta _2}\frac{B}{|v_1-v^\prime _1|} | E(t(v^\prime _1-v_1)+1)|\Big( \int _0^1\sup _{\tau \in [ 0,t]} f^{\#}(\tau ,y,v)dy\Big)  \sup _{(\tau ,X)\in [ 0,t]\times [ 0,1] }f^{\#}(\tau ,X,v_*)dvdv_*d\theta \\
&\leq t(1+\frac{1}{\delta _2})\int B\Big( \int _0^1\sup _{\tau \in [ 0,t]} f^{\#}(\tau ,y,v)dy\Big)  \sup _{(\tau ,X)\in [ 0,t]\times [ 0,1] }f^{\#}(\tau ,X,v_*)dvdv_*d\theta \\
&\leq B_0\pi t(1+\frac{1}{\delta_2})\int \sup _{\tau \in [ 0,t]} f^{\#}(\tau ,y,v)dydv\int \sup _{(\tau ,X)\in [ 0,t]\times [ 0,1] }f^{\#}(\tau ,X,v_*)dv_*.
\end{aligned}\]
Apply Lemma \ref{integral-dxdv}, so that
\begin{equation}\label{bdd-A1}
\int \sup _{x\in [ 0,1] }A_1dv\leq B_0\pi  t(1+\frac{1}{\delta_2}){(c_1^\prime+c'_2)}\int \sup _{(\tau ,X)\in [ 0,t]\times [ 0,1] }f^{\#}(\tau ,X,v_*)dv_*.
\end{equation}
Moreover, performing the change of variables {$(v,v_*,n)\rightarrow (v^\prime _*,v^\prime ,-n)$,}
\begin{eqnarray*}
\int \sup _{x\in [ 0,1] }A_2dv\leq \frac{B_0\pi \sqrt{2}}{\gamma \gamma ^\prime }\sup _{x\in [ 0,1] }\Big( \int _{|y-x|<\delta _2}\sup _{\tau \in [ 0,t]} f^{\#}(\tau ,y,v_*)dydv_*\Big)  \int \sup _{(\tau ,X)\in [ 0,t]\times [ 0,1] }f^{\#}(\tau ,X,v)dv.
\end{eqnarray*}
Given $\delta_1= \frac{\gamma \gamma ^\prime }{4B_0\pi \sqrt{2}}$, apply Lemma \ref{on-small-sets} with the corresponding $\delta_2$ and $t_0$, so that for {$t\leq \min\{T,t_0\}$},
\begin{equation}\label{bdd-A2}
\int \sup _{x\in [ 0,1] }A_2dv\leq \frac{1}{4}\int  \sup _{(\tau ,X)\in [ 0,\textcolor{blue}{t}]\times [ 0,1] }f^{\#}(\tau ,X,v)dv.
\end{equation}
The terms $A_3$ and $A_4$ are treated similarly, with the change of variables $ s\rightarrow y= x+s(v_1-v'_{*1})$. \\
Using (\ref{bdd-A1})-(\ref{bdd-A2}) and the corresponding bounds obtained for $A_3$ and $A_4$ leads to
\[ \begin{aligned}
\int \sup_{(s,x)\in [ 0,t ]\times [0,1]}f^{\#}(s,x,v)dv&\leq 2\int \sup_{x\in [ 0,1] } f_0(x,v)dv\\
&+{4}B_0\pi  t (1+\frac{1}{\delta_2})(c^\prime _1+{c'_2)\int\sup_{(s,x)\in [ 0,t] \times [0,1]} f^{\#}(s,x,v)dv,\quad t\leq \min\{T,t_0\}}.
\end{aligned}\]
Hence
\begin{eqnarray*}
\int \sup_{(s,x)\in [ 0,t ]\times [0,1]}f^{\#}(s,x,v)dv\leq 4\int \sup_{x\in [ 0,1] } f_0(x,v)dv,\quad {t\leq\min \{ t_0, \frac{\delta _2}{8B_0\pi (\delta _2+1)(c^\prime _1+c'_2)} \} . }
\end{eqnarray*}
Since $t_0$, $c'_1$ and $c'_2$ are independent of $\alpha\leq 2^{-L-1}$ and only depend on $\int f_0(x,v)dxdv$, $\int |v|^2f_0(x,v)dxdv$ and $L$, it follows that the argument can be repeated up to $t=T_\alpha$ {with the number of steps uniformly bounded with respect to  $\alpha\leq 2^{-L-1}$}. This completes the proof of the lemma.                       \cqfd
\\
\\
We now prove that the positive time $T_\alpha$ used above, such that $f_\alpha (t)\leq 2^{L+1}$ for $t\in [0,T_\alpha]$, can be taken independent of $\alpha$.
%
% Lemma 3.4
%
\begin{lemma}\label{T-independent-on-alpha}
\hspace{.1cm}\\
Given $f_0\leq2^L$ and satisfying (\ref{hyp-f0}), there is $T\in ] 0,1] $ so that for all $\alpha \in ] 0,2^{-L-1}[ $, the solution $f_\alpha$ to (\ref{f-alpha}) is bounded by $2^{L+1}$ on $[0,T]$.
\end{lemma}
\underline{Proof of Lemma \ref{T-independent-on-alpha}.}\\
Given $\alpha\leq 2^{-L-1}$, it follows from Lemma \ref{T-dependent-on-alpha} that the maximum time $T^\prime _\alpha $ for which $f_\alpha \leq 2^{L+1}$ on $[0,T^\prime _\alpha ]$ is positive. By (\ref{control-by-gain}),
\begin{align*}
&\sup_{s\leq t}f_\alpha ^{\sharp} (s,x,v)
\leq  f_0(x,v)+\int_0^tQ_{\alpha }^{+}(f_\alpha )(s,x+sv_1,v)ds= f_0(x,v)\\
&+\int_0^t\int Bf_\alpha (s,x+sv_1,v')f_\alpha (s,x+sv_1,v'_*)
F_\alpha (f_\alpha )(s,x+sv_1,v)F_\alpha (f_\alpha )(s,x+sv_1,v_*)dv_*d\theta ds.
\end{align*}
With the angular cut-off (2.2), $v_*  \rightarrow v^\prime $ and $v_*  \rightarrow v^\prime _*$ are changes of variables, and so using Lemma \ref{control-mass-density}, the functions $f_\alpha $ for $\alpha \in ] 0,2^{-L-1}[ $satisfy
\begin{align*}
\sup_{(s,x)\in [0,t] \times [0,1] }f_{\alpha }^{\sharp} (s,x,v)&\leq f_0(x,v)+ cB_02^{3L}t\int \sup_{(s,x)\in [0,t] \times [0,1] }f_{\alpha }(s,x,v')dv'\\
&\leq 2^L+cB_02^{3L}tc_1\\
&\leq  3(2^{L-1})\hspace*{0.8in} , t\in [ 0, \min \{ T^\prime _\alpha ,\frac{1}{cc_1B_02^{L-1}} \} ] \, .
\end{align*}
For all $\alpha\leq 2^{-L-1}$, it holds that $T^\prime _\alpha \geq \frac{1}{cc_1B_02^{L-1}} $, else $T^\prime _\alpha $ would not be the maximum time such that $f_\alpha (t)\leq 2^{L+1}$ on $[ 0,T^\prime _\alpha ]$ . Denote by $T= \min \{ 1, \frac{1}{cc_1B_02^{L-1}} \}$. The lemma follows since $T$ does not depend on $\alpha $.\cqfd\\
\\
\\
%
%
%  SECTION 4 PROOF OF THEOREM 1.1
%
%
\section{Proof of Theorem 1.1.}
After the above preparations we
can now prove Theorem \ref{main-theorem}. The conservations of mass, momentum and energy will be proven in Section 5. \\
\hspace*{0.1in}\\
\underline{Proof of Theorem \ref{main-theorem}.}\\
Let us first prove that $(f_\alpha )$ is a Cauchy sequence in $C([0,T]; L^1([0,1] \times \R ^2))$ {with $T$ of Lemma \ref{T-independent-on-alpha}.
For any $(\alpha _1,\alpha _2)\in ]0,1[ ^2$, the function $g=f_{\alpha_1}-f_{\alpha_2}$ satisfies the equation
\[ \begin{aligned}
\partial _tg+v_1\partial _xg&= \int  B(f_{\alpha_1}^{\prime} f_{\alpha_1*}^{\prime } -f_{\alpha_2}^{\prime} f_{\alpha_2 *}^{\prime } )F_{\alpha_1}(f_{\alpha_1})F_{\alpha_1}(f_{\alpha_1 *})dv_*d\theta \\
&-\int  B(f_{\alpha_1}f_{\alpha_1 *} -f_{\alpha _2}f_{\alpha_2 *})F_{\alpha_1}(f_{\alpha_1}^{\prime } )F_{\alpha_1}(f_{\alpha_1 *}^{\prime } )dv_*d\theta \\
&+\int Bf_{\alpha_2}^{\prime } f_{\alpha_2 *}^{\prime } \Big( F_{\alpha_1}(f_{\alpha_1 *})\big( F_{\alpha_1}(f_{\alpha_1})-F_{\alpha_1}(f_{\alpha_2})\big) +F_{\alpha_2}(f_{\alpha_2})\big( F_{\alpha_1}(f_{\alpha_1 *})-F_{\alpha_1}(f_{\alpha_2 *})\big) \Big) dv_*d\theta \\
&+\int Bf_{\alpha_2}^{\prime } f_{\alpha_2 *}^{\prime} \Big( F_{\alpha_1}(f_{\alpha_1 *})\big( F_{\alpha_1}(f_{\alpha_2})-F_{\alpha_2}(f_{\alpha_2})\big) +F_{\alpha_2}(f_{\alpha_2})\big( F_{\alpha_1}(f_{\alpha_2 *})-F_{\alpha_2}(f_{\alpha_2 *})\big) \Big) dv_*d\theta \\
& -\int Bf_{\alpha_2}f_{\alpha_2 *}\Big( F_{\alpha_1}(f_{\alpha_1 *}^{\prime} )\big( F_{\alpha_1}(f_{\alpha_1}^{\prime } )-F_{\alpha_1}(f_{\alpha_2}^{\prime } )\big) +F_{\alpha_2}(f_{\alpha_2}^{\prime} )\big( F_{\alpha_1}(f_{\alpha_1*}^{\prime } )-F_{\alpha_1}(f_{\alpha_2 *}^{\prime } )\big) \Big) dv_*d\theta \\
&-\int Bf_{\alpha_2}f_{\alpha_2 *}\Big( F_{\alpha_1}(f_{\alpha_1*}^{\prime} )\big( F_{\alpha_1}(f_{\alpha_2}^{\prime } )-F_{\alpha_2}(f_{\alpha_2}^{\prime} )\big) +F_{\alpha_2}(f_{\alpha_2}^{\prime} )\big( F_{\alpha_1}(f_{\alpha_2 *}^{\prime} )-F_{\alpha_2}(f_{\alpha_2 *}^{\prime} )\big) \Big) dv_*d\theta .\hspace{.1cm} (4.8)
\end{aligned}\]
Using Lemma \ref{control-mass-density} and taking $\alpha_1,\alpha_2< 2^{-L-1}$,
\[ \begin{aligned}
\int  B&\Big(\lvert f_{\alpha_1}f_{\alpha_1*} -f_{\alpha_2}f_{\alpha_2 *}\rvert F_{\alpha_1}(f_{\alpha_1}^{\prime} )F_{\alpha_1}(f_{\alpha_1*}^{\prime} )\Big)^\sharp dxdvdv_*d\theta \\
&\leq c2^{2L}\Big( \int \sup _{x\in [ 0,1] }f_{\alpha_1}^{\sharp} (t,x,v)dv
+ \int \sup_ {x\in [ 0,1] }f_{\alpha_2}^{\sharp} (t,x,v)dv\Big) \int \lvert (f_{\alpha_1} -f_{\alpha_2} )^{\sharp }(t,x,v)\rvert dxdv\\
&\leq cc_12^{2L}\int \lvert g^\sharp(t,x,v)\rvert dxdv.
\end{aligned}\]
We similarly obtain
\begin{eqnarray*}
\int B \Big( f_{\alpha_2}^{\prime} f_{\alpha_2 *}^{\prime} F_{\alpha_1}(f_{\alpha_1 *})\lvert ( F_{\alpha_1}(f_{\alpha_2})-F_{\alpha_2}(f_{\alpha_2})\lvert )\Big) ^\sharp dxdvdv_*d\theta\leq cc_12^{2L}|\alpha_1-\alpha_2| ,
\end{eqnarray*}
and
\begin{eqnarray*}
\int B\Big(f_{\alpha_2}f_{\alpha_2 *} F_{\alpha_1}(f_{\alpha_1 *}^{\prime} )\lvert F_{\alpha_1}(f_{\alpha_1}^{^\prime} )-F_{\alpha_1}(f_{\alpha_2}^{\prime} )\rvert \Big)^\sharp dxdvdv_*d\theta\leq cc_12^L\int|g^\sharp(t,x,v)|dxdv.
\end{eqnarray*}
The remaining terms are estimated in the same way. It follows
\begin{eqnarray*}
\frac{d}{dt}\int|g^\sharp(t,x,v)|dxdv\leq cc_12^{2L}\Big(\int|g^\sharp (t,x,v)|dxdv+|\alpha_1-\alpha_2|\Big).
\end{eqnarray*}
 Hence
 \begin{eqnarray*}
 \lim _{(\alpha _1,\alpha _2)\rightarrow (0,0)}\sup _{t\in [0,T]}\int |g^\sharp(t,x,v)|dxdv= 0.
 \end{eqnarray*}
 And so $(f_\alpha )$ is a Cauchy sequence in $C([0,T]; L^1([0,1] \times \R ^2))$. Denote by $f$ its limit. With analogous arguments to the previous ones in the proof of this lemma, it holds that
 \begin{eqnarray*}
 \lim _{\alpha \rightarrow 0}\int \lvert Q(f)-Q(f_\alpha )\rvert (t,x,v)dtdxdv= 0.
 \end{eqnarray*}
 Hence $f$ is a strong solution to (\ref{f}) on $[0,T]$ with initial value $f_0$. If there were two solutions, their difference denoted by $G$ would with similar arguments satisfy
 \begin{eqnarray*}
\frac{d}{dt}\int |G^\sharp(t,x,v)|dxdv\leq cc_12^{2L}\int |G^\sharp(t.x.v)|dxdv,
 \end{eqnarray*}
 hence be identically equal to its initial value zero. Finally, if $f_1$ (resp. $f_2$) is the solution to (\ref{f}) with initial value $f_{10}$ (resp. $f_{20}$), then similar arguments lead to
\begin{eqnarray*}
\frac{d}{dt}\int |(f_1-f_2)^\sharp(t,x,v)|dxdv\leq cc_12^{2L}\int |(f_1-f_2)^\sharp(t,x,v)|dxdv,
 \end{eqnarray*}
 so that
 \begin{eqnarray*}
\parallel (f_1-f_2)(t,\cdot ,\cdot )\parallel _{L^1([0,1] \times \R ^2)}\leq e^{cc_12^{2L}T}\parallel f_{10}-f_{20}\parallel _{L^1([0,1] \times \R ^2)},\quad t\in [0,T] ,
 \end{eqnarray*}
i.e. stability holds. If $\sup_{(x,v)\in[ 0,1] \times \R ^2}f(T,x,v)<2^{L+1}$, then the procedure can be repeated, i.e. the same proof can be carried out from the initial value $f(T)$. This leads to a maximal interval denoted by $[0,\tilde {T}_1]$ on which $f(t,\cdot ,\cdot )\leq 2^{L+1}$. By induction there exists an increasing  sequence of times $(\tilde{T}_n)$  such that $f(t,\cdot ,\cdot )\leq 2^{L+n}$ on $[ 0,\tilde{T}_n]$. Let $T_\infty = \lim _{n\rightarrow +\infty }\tilde{T}_n$. Either $\tilde{T}_\infty = +\infty $ and the solution $f$ is global in time, or $T_\infty $ is finite and the solution tends to infinity in the $L^\infty$-norm at $T_\infty$. \cqfd
 %
 %
%
% SECTION 5: CONSERVATIONS OF MASS, MOMENTUM AND ENERGY
%
\section{Conservations of mass, momentum and energy.}
\setcounter{equation}{0}
\setcounter{theorem}{0}
The following  two preliminary lemmas are needed for the control of large velocities.
%
% Lemma 5.1 (former Lemma 3.6)
%
\begin{lemma}\label{large-velocities1}
\hspace*{0.02in}\\
The solution $f$ of (\ref{f}) with initial value $f_0$, satisfies
\begin{eqnarray*}
\int _0^1\int_{|v|>\lambda} |v|\sup_{t\in [ 0,T] }f^\sharp (t,x,v)dvdx\leq\frac{c_T}{\lambda},\quad t\in [0,T],
\end{eqnarray*}
where $c_T$ only depends on $T$, {$\int f_0(x,v)dxdv$ and $\int |v|^2f_0(x,v)dxdv$.}
\end{lemma}
\underline{Proof of Lemma \ref{large-velocities1}.} \\
As in (\ref{control-by-gain}),
\begin{eqnarray*}
 \sup_{t\in [ 0,T] }f^\sharp (t,x,v)
\leq  f_0(x,v)+\int_0^TQ^+(f)(s,x+sv_1,v)ds.
\end{eqnarray*}
Integration with respect to $(x,v)$  for $|v|>\lambda$, gives
\begin{eqnarray*}
 \int _0^1\int_{|v|>\lambda}|v|\sup_{t\in [ 0,T] }f^\sharp (t,x,v)dvdx
\leq  \int \int_{|v|>\lambda}|v|f_0(x,v)dvdx+\int_0^T\int_{|v|>\lambda} B\\
|v| f(s,x+sv_1,v^\prime )f(s,x+sv_1,v^\prime _*)F(f)(s,x+sv_1,v)F(f)(s,x+sv_1,v_*)dvdv_*d\theta dxds.
\end{eqnarray*}
Here in the last integral, either $|v^\prime |$ or $|v^\prime _*|$ is the largest and larger than $\frac{\lambda}{\sqrt 2}$. The two cases are symmetric, and we discuss the case $|v^\prime |\geq|v^\prime _*|$. After a translation in $x$, the integrand is estimated {from above} by
\begin{eqnarray*}
c |v^\prime |f^{\#} (s,x,v^\prime )\sup_{(t,x)\in [ 0,T] \times [0,1]}f^{\#}(t,x,v^\prime _*).
\end{eqnarray*}
 The change of variables {$(v,v_*,n)\rightarrow (v^\prime ,v^\prime _*,-n)$}, the integration over
 \begin{eqnarray*}
(s,x,v,v_*,\omega )\in [ 0,T] \times [ 0,1] \times \{ v\in \R ^2; |v| >\frac{\lambda}{\sqrt 2}\} \times \R ^2 \times [ -\frac{\pi }{2}, \frac{\pi }{2}] ,
\end{eqnarray*}
and Lemma \ref{control-mass-density} give the bound
\begin{eqnarray*}
\frac{c}{\lambda}\Big( \int_0^T\int |v|^2f^{\#}(s,x,v)dxdvds\Big) \Big( \int \sup_{(t,x)\in [ 0,T] \times [0,1]} f^{\#}(t,x,v_*)dv_*\Big) \leq \frac{cTc_1(T)}{\lambda}\int |v|^2f_0(x,v)dxdv.
\end{eqnarray*}
The lemma follows.                        \cqfd
%
%
% Lemma 5.2 (former Lemma 3.7)
%
\begin{lemma}\label{large-velocities2}
\hspace*{0.02in}\\
The solution $f$ of (\ref{f}) with initial val;ue $f_0$ satisfies
\begin{eqnarray*}
\int_{|v|>\lambda} \sup_{(t,x)\in [ 0,T] \times [0,1]}f^\sharp (t,x,v)dv\leq\frac{c^\prime _T}{\sqrt{\lambda}},\quad t\in [0,T],
\end{eqnarray*}
where $c^\prime _T$ only depends on $T$, {$\int f_0(x,v)dxdv$ and $\int |v|^2f_0(x,v)dxdv$.}
\end{lemma}
\underline{Proof of Lemma \ref{large-velocities2}.}\\
Take $\lambda >2$. As above,
\begin{eqnarray}
 \int _{|v|>\lambda }\sup_{(t,x)\in [ 0,T] \times [ 0,1] }f^\sharp (t,x,v)dv
\leq  \int _{|v|>\lambda }\sup_{x\in [ 0,1] }f_0(x,v)dv+cC,
\end{eqnarray}
where
\begin{eqnarray*}
C= \int _{|v|>\lambda }\sup _{x\in [ 0,1] }\int_0^T\int B
 f^{\#}(s,x+s(v_1-v'_1),v')f^{\#}(s,x+s(v_1-v'_{*1}),v'_*)dvdv_*d\theta ds.
\end{eqnarray*}
For $v',v'_*$ outside of the angular cutoff {(\ref{hyp2-B})}, let $n$ be the unit vector in the direction $v-v'$, and $n_{\perp}$ the orthogonal
unit vector in the direction $v-v'_*$. Let $e_1$ be a unit vector in the $x$-direction.\\
Split $C$ as $C= \sum _{1\leq i\leq 6}C_i$, where $C_1$ (resp. $C_2$, $C_3$) refers to integration with respect to $(v_*, \theta )$ on
\begin{eqnarray*}
\{ (v_*,\theta ); \quad n\cdot e_1\geq \frac{1}{\sqrt{2}}, \quad  |v'| \geq |v'_*|\} ,
\end{eqnarray*}
\begin{eqnarray*}
\big( \text{resp.    }\{ (v_*,\theta ); n\cdot e_1\geq \sqrt{1-\frac{1}{\lambda }}, \hspace*{0.03in}|v'| \leq |v'_*|\} ,\quad \{ (v_*,\theta ); n\cdot e_1\in [\frac{1}{\sqrt{2}}, \sqrt{1-\frac{1}{\lambda }}], \hspace*{0.03in}|v'| \leq |v'_*|\}\big) ,
\end{eqnarray*}
and analogously for $C_i$, $4\leq i\leq 6$, with $n$ replaced by $n_\perp $.
By symmetry, $C_i$, $4\leq i\leq 6$ can be treated as $C_i$, $1\leq i\leq 3$, so we only discuss the control of $C_i$, $1\leq i\leq 3$.\\
By the change of variables {$(v,v_*,n)\rightarrow (v^\prime ,v^\prime _*,-n)$,} and noticing that $ |v'| \geq \frac{\lambda }{\sqrt{2}}$ in the domain of integration of $C_1$, it holds that
\[ \begin{aligned}
C_1&\leq \int _{|v|>\frac{\lambda }{\sqrt{2}}}\sup _{x\in [ 0,1] }\int_0^T\int _{n\cdot e_1\geq \frac{1}{\sqrt{2}}}B
 f^{\#}(s,x+s(v^\prime _1-v_1),v)f^{\#}(s,x+s(v^\prime _1-v_{*1}),v_*)dv_*d\theta dsdv\\
& \leq \int _{|v|>\frac{\lambda }{\sqrt{2}}}\sup _{x\in [ 0,1] }\int_0^T\int _{n\cdot e_1\geq \frac{1}{\sqrt{2}}}B
 \sup _{\tau \in [ 0,T] }f^{\#}(\tau ,x+s(v^\prime _1-v_1),v)\sup _{(\tau ,X)\in [ 0,T] \times [ 0,1] }f^{\#}(\tau ,X,v_*)dv_*d\theta dsdv.
\end{aligned}\]
With the change of variables $s\rightarrow y= x+s(v^\prime _1-v_1)$,
\[ \begin{aligned}
C_1&\leq \int _{|v|>\frac{\lambda }{\sqrt{2}}}\sup _{x\in [ 0,1] }\int _{n\cdot e_1\geq \frac{1}{\sqrt{2}}}\int _{y\in (x,x+T(v^\prime _1-v_1))}\frac{B}{|v^\prime _1-v_1|}
  \sup _{\tau \in [ 0,T] }f^{\#}(\tau ,y,v)\sup _{(\tau ,X)\in [ 0,T] \times [ 0,1] }f^{\#}(\tau ,X,v_*)dydv_*d\theta dv\\
   &\leq  \int _{|v|>\frac{\lambda }{\sqrt{2}}}\int _{n\cdot e_1\geq \frac{1}{\sqrt{2}}} \frac{{ |E(T(v^\prime _1-v_1))+1) |}}{|v^\prime _1-v_1|}\int _0^1B
  \sup _{\tau \in [ 0,T] }f^{\#}(\tau ,y,v)\sup _{(\tau ,X)\in [ 0,T] \times [ 0,1] }f^{\#}(\tau ,X,v_*)dydv_*d\theta dv.
  \end{aligned}\]
  {Moreover,
  \begin{eqnarray*}
  |E(T(v^\prime _1-v_1))+1) | \leq T | v_1^\prime -v_1| +{1}\leq \big( T+\frac{{\sqrt{2}}}{\gamma \gamma ^\prime }\big) | v_1^\prime -v_1| ,
  \end{eqnarray*}
where $\gamma $ and $\gamma ^\prime $ were defined in (2.2). Consequently, }
  \[ \begin{aligned}
  C_1&\leq  c(T+1)\int _0^1\int _{|v|>\frac{\lambda }{\sqrt{2}}}
  \sup _{\tau \in [ 0,T] }f^{\#}(\tau ,y,v)dydv\int \sup _{(\tau ,X)\in [ 0,T] \times [ 0,1] }f^{\#}(\tau ,X,v_*)dv_*\\
 &\leq  \frac{c(T+1)}{\lambda }\int _0^1\int _{|v|>\frac{\lambda }{\sqrt{2}}}
  |v|\sup _{\tau \in [ 0,T] }f^{\#}(\tau ,y,v)dydv\int \sup _{(\tau ,X)\in [ 0,T] \times [ 0,1] }f^{\#}(\tau ,X,v_*)dv_*.
\end{aligned}\]
By Lemmas \ref{control-mass-density} and \ref{large-velocities1},
\begin{eqnarray*}
C_1\leq \frac{c}{\lambda ^2}(T+1)c_T c_1(T).
\end{eqnarray*}
Moreover,
\[ \begin{aligned}
C_2 &\leq \int _{|v^\prime |>\lambda , |v_*|> |v|, n\cdot e_1\geq \sqrt{1-\frac{1}{\lambda }}}\frac{B}{|v^\prime _1-v_1|}\\
&\sup _{x\in [ 0,1] }\int_{y\in (x,x+T(v^\prime _1-v_1))}
 \sup _{\tau \in [ 0,T] }f^{\#}(\tau ,y,v)
\sup _{(\tau ,X)\in [ 0,T] \times [ 0,1] }f^{\#}(\tau ,X,v_*)dydvdv_*d\theta \\
&\leq  c(T+1)\int _{n\cdot e_1\geq \sqrt{1-\frac{1}{\lambda }}}d\theta \int
 \sup _{\tau \in [ 0,T] }f^{\#}(\tau ,y,v)dydv\int \sup _{(\tau ,X)\in [ 0,T] \times [ 0,1] }f^{\#}(\tau ,X,v_*)dv_*\\
& \leq \frac{c}{\sqrt{\lambda }}(T+1)^2c_1(T),
\end{aligned}\]
by Lemmas \ref{integral-dxdv} and \ref{control-mass-density}.
Finally,
\[ \begin{aligned}
C_3& \leq \int _{|v_*|>\frac{\lambda }{\sqrt{2}}, \frac{1}{\sqrt{\lambda }}\leq n_\perp \cdot e_1\leq \frac{1}{\sqrt{2}}}\sup _{(\tau ,X)\in [ 0,T] \times [ 0,1] }f^{\#}(\tau ,X,v)\frac{B}{|v^\prime _1-v_{*1}|}\\
&\hspace*{0.3in}\sup _{x\in [ 0,1] }\Big( \int_{y\in (x,x+T(v^\prime _1-v_{*1}))}
 \sup _{\tau \in [ 0,T] }f^{\#}(\tau ,y,v_*)dy\Big) dvdv_*d\theta \\
&\leq  c(T+1)\sqrt{\lambda}\Big( \int \sup _{(\tau ,X)\in [ 0,T] \times [ 0,1] }f^{\#}(\tau ,X,v)dv\Big) \Big( \int_{|v_*|>\frac{\lambda }{\sqrt{2}}} \sup _{\tau \in [ 0,T] }f^{\#}(\tau ,y,v_*)dydv_*\Big) .
\end{aligned}\]
By Lemmas \ref{control-mass-density} and \ref{large-velocities1},
\[ \begin{aligned}
C_3&\leq  {\frac{c}{\sqrt{\lambda}}(T+1)c_1(T)c_T.}
\end{aligned}\]
\hspace*{0.1in}\\
The lemma follows. \cqfd
%
% Lemma 5.3
%
\begin{lemma}\label{conservations}
The solution $f$ to (\ref{f}) with initial value $f_0$ conserves mass, momentum and energy.
\end{lemma}
\underline{Proof of Lemma  \ref{conservations}.}\\
The conservation of mass and first momentum of $f$ will follow from the boundedness of the total energy. The energy is non-increasing since the approximations $f_\alpha $ conserve energy and
\begin{eqnarray*}
\lim _{\alpha \rightarrow 0}\int _0^1\int _{\lvert v\rvert <V}\lvert (f-f_\alpha )(t,x,v)\rvert \lvert v\rvert ^2dxdv= 0,\quad \text{for all }t\in [0,T] \text{   and positive  }V.
\end{eqnarray*}
 Energy conservation will be satisfied if the energy is non-decreasing. Taking $\psi_\epsilon=\frac{|v^2|}{1+\epsilon|v|^2}$  as approximation for $|v|^2$, it is enough to bound
\begin{eqnarray*}
\int Q(f)(t,x,v)\psi_\epsilon (v)dxdv = \int B\psi_{\epsilon}\Big( f^\prime f^\prime _{*}F(f)F(f_{*})
- ff_{*}F(f^\prime )F(f^\prime _{*})\Big) dxdvdv_*d\theta
\end{eqnarray*}
from below by zero in the limit $\epsilon \rightarrow 0$. Similarly to{ \cite{Lu2}},
\[ \begin{aligned}
\int Q(f)\psi_\epsilon dxdv
&=\frac{1}{2}\int B ff_{*}F(f^\prime )F(f^\prime _{*}\Big( \psi_\epsilon(v')+\psi_\epsilon(v'_*)-\psi_\epsilon(v)-\psi_\epsilon(v_*)
\Big)dxdvdv_*d\theta \\
&\geq -\int Bff_{*}F(f^\prime )F(f^\prime _{*})\frac{\epsilon |v|^2|v_*|^2}{(1+\epsilon|v|^2)(1+\epsilon|v_*|^2)}dxdvdv_*d\theta .
\end{aligned}\]
The previous line, with the integral taken over a bounded set in $(v,v_*)$, converges to zero when $\epsilon\rightarrow 0$. In  integrating over $|v|^2+|v_*|^2\geq2\lambda^2$ , there is symmetry between the subset of the domain with $|v|^2>\lambda^2$ and the one with $|v_*|^2>\lambda^2$. We discuss the first sub-domain, for which the integral in the last line is bounded from below by
\[ \begin{aligned}
&-c\int |v_*|^2f(t,x,v_*)dxdv_*\int_{|v|\geq \lambda} B \sup_{(s,x)\in [ 0,t] \times [0,1]}f^\#(s,x,v)dvd\theta\\
&\geq -c\int_{|v|\geq \lambda} \sup_{0\leq (s,x)\in [ 0,t] \times [0,1]}f^\#(s,x,v)dv.
\end{aligned}\]
It follows from Lemma \ref{large-velocities2} that the right hand side tends to zero when $\lambda \rightarrow \infty$.
This implies that the energy is non-decreasing, and bounded from below by its initial value. That completes the proof of the lemma.     \cqfd
\\
\\
\\

\[\]

\end{document}